\documentclass[times,twocolumn,review]{elsarticle}

\usepackage{medima}
\usepackage{framed,multirow}

\usepackage{amssymb}
\usepackage{latexsym}
\usepackage{url}
\usepackage{xcolor}
\usepackage{amsmath}
\usepackage{graphicx}
\usepackage{booktabs}
\usepackage{hyperref}
\usepackage{pifont}

\definecolor{newcolor}{rgb}{.8,.349,.1}

\journal{Medical Image Analysis}

\begin{document}

\verso{Wenhui Cui \textit{et~al.}}

\begin{frontmatter}

\title{Meta Transfer of Self-Supervised Knowledge: Foundation Model in Action for Post-Traumatic Epilepsy Prediction}

\author[1]{Wenhui \snm{Cui}}
\author[1]{Haleh \snm{Akrami}}
\author[1]{Ganning \snm{Zhao}}
\author[1]{Anand A. \snm{ Joshi}}
\author[1]{Richard M. \snm{Leahy}\corref{cor1}}
\cortext[cor1]{Corresponding author at: Ming Hsieh Department of Electrical and Computer Engineering, 
      University of Southern California, Los Angeles 90089, United States. 
  Email address: leahy@usc.edu}

\address[1]{Ming Hsieh Department of Electrical and Computer Engineering, 
      University of Southern California, Los Angeles 90089, United States}


\begin{abstract}
Despite the impressive advancements achieved using deep-learning for functional brain activity analysis, the heterogeneity of functional patterns and scarcity of imaging data still pose challenges in tasks such as prediction of future onset of Post-Traumatic Epilepsy (PTE) from data acquired shortly after traumatic brain injury (TBI). Foundation models pre-trained on separate large-scale datasets can improve the performance from scarce and heterogeneous datasets. For functional Magnetic Resonance Imaging (fMRI), while data may be abundantly available from healthy controls, clinical data is often scarce, limiting the ability of foundation models to identify clinically-relevant features. We overcome this limitation by introducing a novel training strategy for our foundation model by integrating meta-learning with self-supervised learning to improve the generalization from normal to clinical features. In this way we enable generalization to other downstream clinical tasks, in our case prediction of PTE. To achieve this, we perform self-supervised training on the control dataset to focus on inherent features that are not limited to a particular supervised task while applying meta-learning, which strongly improves the model's generalizability using bi-level optimization.  
Through experiments on neurological disorder classification tasks, we demonstrate that the proposed strategy significantly improves task performance on small-scale clinical datasets. 
To explore the generalizability of the foundation model in downstream applications, we then apply the model to an unseen TBI dataset for prediction of PTE using zero-shot learning. Results further demonstrated the enhanced generalizability of our foundation model.
\end{abstract}

\begin{keyword}
\KWD Foundation model \sep Meta learning\sep Self-supervised learning\sep Generalization \sep fMRI
\end{keyword}

\end{frontmatter}


\section{Introduction}
Deep learning based approaches have demonstrated success in analyzing brain connectivity based on functional magnetic resonance imaging (fMRI)~\cite{gadgil2020spatio,ahmedt2021graph}, but the scarcity and heterogeneity of fMRI data still pose challenges in clinical applications such as predicting the future onset of Post-Traumatic Epilepsy (PTE) from acute data acquired shortly after traumatic brain injury (TBI)~\cite{akbar2022post}. 
Identification of subjects at high risk of developing PTE can eliminate the need to wait for spontaneous epileptic seizures to occur before starting treatment and enable the mitigation of risks to subjects whose
seizures could result in serious injury or death. fMRI plays a vital role in identifying biomarkers for PTE. The presence of lesions in TBI patients can alter resting-state brain dynamics~\cite{palacios2013resting}. This will be reflected in fMRI data collected after injury, which can therefore provide valuable biomarkers for PTE. However, 
TBI datasets are usually characterized by high variability among subjects and limited numbers of subjects, presenting a significant challenge for training of deep learning methods to predict PTE. 
To tackle these challenges, we can employ a large pre-trained model. 
Developing a foundation model pre-trained on large-scale datasets has been exceptionally successful in natural language processing~\cite{radford2019language} and computer vision tasks~\cite{yu2023language}. 
Typically, foundation models can generalize across domains and tasks, achieving promising performance even in few-shot and zero-shot learning scenarios. Foundation models are usually trained using a self-supervised task~\cite{radford2019language} involving extensive and diverse datasets. In medical data, it is common to have a large amount of healthy control data, while simultaneously facing a scarcity of clinical data collected for any particular neurological disorder. Simply aggregating all normal and clinical data and applying self-supervised learning may cause limited generalization and bias because of data imbalance and heterogeneity in clinical features. This can in-turn lead to poor performance in the group with clinical pathology~\cite{azizi2023robust}. Since our goal is to achieve superior performance on downstream clinical tasks, it is crucial to learn how to generalize to useful clinical features during the training of the foundation model.

To address the limited generalization, we adopt meta-learning as a novel approach for developing foundation models that leverage features from large-scale normal datasets and small-scale clinical datasets. Meta-learning has recently gained tremendous attention because of its learning-to-learn mechanism, which strongly increases the generalizability of models across different tasks~\cite{zhang2019metapred, liu2020meta, finn2017model} and has shown success in few-shot learning tasks. Meta-learning enhances the model's generalization, even when trained on smaller-scale datasets. 
 One of the most popular meta-learning algorithms, the Model Agnostic Meta-Learning method (MAML)~\cite{finn2017model}, is a gradient-based approach that uses a bi-level optimization scheme to enable the model to learn how to generalize on an unseen domain during training. However, in the context of fMRI, the availability of diverse datasets is typically limited. 
Instead of learning to generalize from multiple source tasks to multiple target tasks in MAML,~\cite{liu2020meta} propose a meta representation learning approach to learn generalizable features from one source domain and improve the generalization to one target domain. 
To combine data acquired from healthy (control) populations with clinical data from patients, we consider a source domain with abundant control data and a target domain with limited clinical data during upstream training of the foundation model. Through meta-learning~\cite{liu2020meta}, the model is enabled to generalize from control features to clinical features. For downstream applications, we focus on a TBI/PTE dataset characterized by extreme heterogeneity and scarcity of clinical fMRIs, where traditional deep learning models often over-fit and fail to generalize. Our goal is to apply the meta-learning pre-trained model to this PTE dataset during downstream adaptation. By leveraging the learned generalization from normal to clinical features, we aim to enhance the model performance on this challenging clinical dataset.
 
Self-supervised learning has shown the ability to improve the generalization of features in foundation models~\cite{ortega2023brainlm, thomas2022self, azizi2023robust}. In contrast to fully-supervised tasks such as classification or segmentation, self-supervised tasks are typically designed to learn intrinsic features that are not specific to a particular task~\cite{taleb20203d}. Contrastive self-supervised learning applied to fMRI classification has demonstrated the ability to prevent over-fitting on small medical datasets and address high intra-class variances~\cite{wang2022contrastive}. 
For our foundation model, we apply contrastive self-supervised learning, known to be effective in representation learning~\cite{azizi2023robust}, to the control data (the source domain in the meta-learning framework) to learn more generalizable features.

We propose a novel training strategy for the foundation model: Meta Transfer of Self-supervised Knowledge (MeTSK), which harnesses meta-learning to facilitate the transfer of self-supervised features from large-scale control to scarce clinical datasets. This training strategy is designed to enhance the foundation model's capacity to generalize from normal features to clinical features, which will also facilitate the generalization to new and unseen clinical features in downstream applications. 
The proposed network architecture consists of a feature extractor that learns general features from both source (control) and target  (clinical) domains, and source and target heads to learn domain-specific features for the source and target domain, respectively. The bi-level optimization strategy is applied to learn generalizable features using a Spatio-temporal Graph Convolutional Network (ST-GCN)~\cite{gadgil2020spatio} as the backbone model. To further validate the generalization improvement achieved by MeTSK, we adopt domain similarity, representing the least amount of work required to transform features into a different domain, as our generalization metric. A larger domain similarity implies that the features are more transferable. Our experimental results demonstrate the effectiveness of MeTSK on neurological disorder classification tasks by improving both inter-domain and intra-domain generalization. This finding underscores the potential of MeTSK in effectively bridging the gap between normal and clinical datasets. 

Beyond the typical approach of fine-tuning the entire foundation model for downstream adaptation, linear probing is a crucial method for evaluating the quality of features learned by the foundation model~\cite{chen2020generative, kumar2022fine}. Linear probing involves freezing the parameters of a pre-trained model and training a linear classifier on the output. The intuition behind linear probing is that good features should be linearly separable between classes~\cite{chen2020generative}. 
For our downstream application, 
we perform linear probing on the PTE prediction task. We apply our foundation model trained using MeTSK to directly generate features for the PTE fMRI data without any fine-tuning. We then input these features to a linear classifier and achieved superior classification performance compared to using functional connectivity features as input. 
In summary, our contribution is two-fold: 
\begin{itemize}
    \item We propose a novel training strategy for developing a foundation model for fMRI data by learning how to generalize from control to clinical features;
    \item We address the heterogeneity and scarcity of clinical fMRI data by improving the generalization of the model through the integration of meta-learning and self-supervised learning.
\end{itemize}

\section{Related Work}
\subsection{Foundation Models for fMRI}
Foundation models pre-trained on large-scale data have shown remarkable performance in tasks including image and video generation~\cite{yu2023language}, speech recognition~\cite{rubenstein2023audiopalm}, and medical question answering~\cite{singhal2023expertlevel}. 
Recently,~\cite{thomas2022self} adapted several prominent models in natural language processing including BERT~\cite{devlin2018bert} and GPT~\cite{radford2019language}, to learn the dynamics of brain activity in fMRIs. The models are trained on massive fMRI data from 11,980 experimental scans of 1,726 individuals across 34 datasets. A self-supervised task is adopted during training. The trained model is then fine-tuned on benchmark mental state decoding
datasets and achieved improvements compared to the same model trained from scratch. 
BrainLM~\cite{ortega2023brainlm} is a recently published foundation model for brain activity dynamics trained on 6,700 hours of fMRI recordings. The model consists of a Transformer-based~\cite{vaswani2017attention} masked auto-encoder architecture adapted from BERT~\cite{devlin2018bert} and Vision Transformer~\cite{dosovitskiy2020image}. During pre-training, BrainLM incorporates a self-supervised task that predicts the masked segments of time series in fMRI data, which is similar to the pre-training task in~\cite{thomas2022self}. They fine-tuned the model to predict metadata variables
acquired from the UK BioBank dataset~\cite{allen2014uk} and achieved superior performance. 
In contrast to previous work that developed foundation models for fMRI using extensive datasets comprising vast fMRI recordings, we propose a training strategy for a foundation model with relatively limited data and focus on improving the generalization of the model to downstream clinical applications.

\subsection{Prediction of Post-Traumatic Epilepsy}
Survivors of Traumatic Brain Injury (TBI) often experience significant disability due to their injuries \cite{parikh2007traumatic}. These injuries can lead to a range of physical and psychological effects, with some symptoms appearing immediately and others developing over time. Post-traumatic epilepsy (PTE) refers to recurrent and unprovoked post-traumatic seizures occurring after 1 week~\cite{verellen2010post}. Identifying individual prognostic markers for PTE is crucial \cite{engel2013epilepsy}, as it can reduce the time and cost for TBI patients to begin clinical trials and decrease the risk of severe injury or death due to seizures.
The prediction and prevention of PTE development remains a significant challenge. Animal studies in adult male Sprague-Dawley rats have shown the potential of MRI-based image analysis in identifying biomarkers for PTE \cite{immonen2013mri,pitkanen2016advances}. These studies indicate the involvement of the perilesional cortex, hippocampus, and temporal lobe in PTE \cite{pitkanen2012head}. Despite progress, brain imaging is still not fully leveraged in PTE biomarker research. Various human neuroimaging studies have provided insights into TBI \cite{dennis2016tensor,farbota2012longitudinal,kim2008structural} and epilepsy \cite{li2009detection,mo2019automated,sollee2022artificial}, but fMRI-based PTE prediction is limited. 

Clinical and research studies in epilepsy often include both anatomical (MRI, CT) and functional (PET, EEG, MEG, ECoG, depth electrodes, fMRI) mapping. While epileptogenic zones can be found in almost any location in the brain, the temporal lobe and the hippocampus are the most common sites causing focal epileptic seizures \cite{sollee2022artificial}. Multimodal MRI and PET imaging has been used to predict the laterality of temporal lobe epilepsy \cite{pustina2015predicting,sollee2022artificial}. Extensive changes in brain networks due to epilepsy were reported using PET, fMRI, and diffusion imaging \cite{li2009detection,pitkanen2016advances,pustina2015predicting,akrami2021prediction,sollee2022artificial}. 
Recent studies employing machine learning to identify potential PTE biomarkers \cite{la2019machine,akrami2021prediction,akrami2022neuroanatomic} have primarily focused on pairwise correlation patterns in resting fMRI signals. However, the heterogeneity of PTE functional activity and data scarcity often lead to over-fitting and limited generalization in deep-learning approaches. Here we similarly focus on the use of only fMRI in PTE prediction, but with the novel use of a foundation-model approach for this problem. 

\section{Methods}
Here we introduce our proposed strategy, MeTSK, which improves the generalization of self-supervised fMRI features from a control dataset to a clinical dataset. Assume there exists a source domain (healthy controls) $\mathcal{S}$ with abundant training data $X_\mathcal{S}$ and a target domain (clinical) $\mathcal{T}$, where the training data $X_\mathcal{T}$ is limited. A feature extractor $f(\phi)$, a target head $h_\mathcal{T}(\theta_t)$, and a source head $h_\mathcal{S}(\theta_s)$ are constructed to learn source features $h_\mathcal{S}(f(X_\mathcal{S}; \phi); \theta_s)$ as well as target features  $h_\mathcal{T}(f(X_\mathcal{T}; \phi); \theta_t)$, where $\phi$, $\theta_t$, and $\theta_s$ are model parameters. 
\begin{figure*}[!t]
    \centering
\includegraphics[scale=.55]{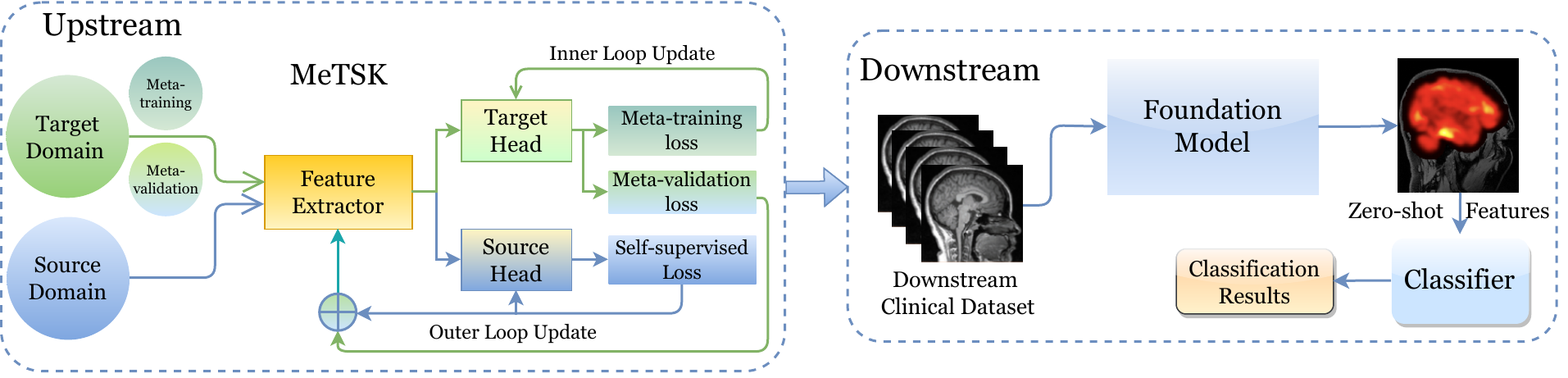}
    \caption{An illustration of the proposed MeTSK strategy for upstream training and downstream applications. In MeTSK, two optimization loops are involved in training. The inner loop only updates the target head, while the outer loop updates the source head and feature extractor. For downstream applications, we directly apply the pre-trained foundation model without any fine-tuning and generate zero-shot features for the downstream dataset. The zero-shot features are then used to train a simple classifier and generate final classification results.
    }
    \label{fig1}
\end{figure*}
The overall framework of MeTSK and the foundation model pipeline is illustrated in Fig.~\ref{fig1}.

\subsection{Feature Extractor: ST-GCN}
\begin{figure*}[!t]
    \centering
\includegraphics[scale=.7]{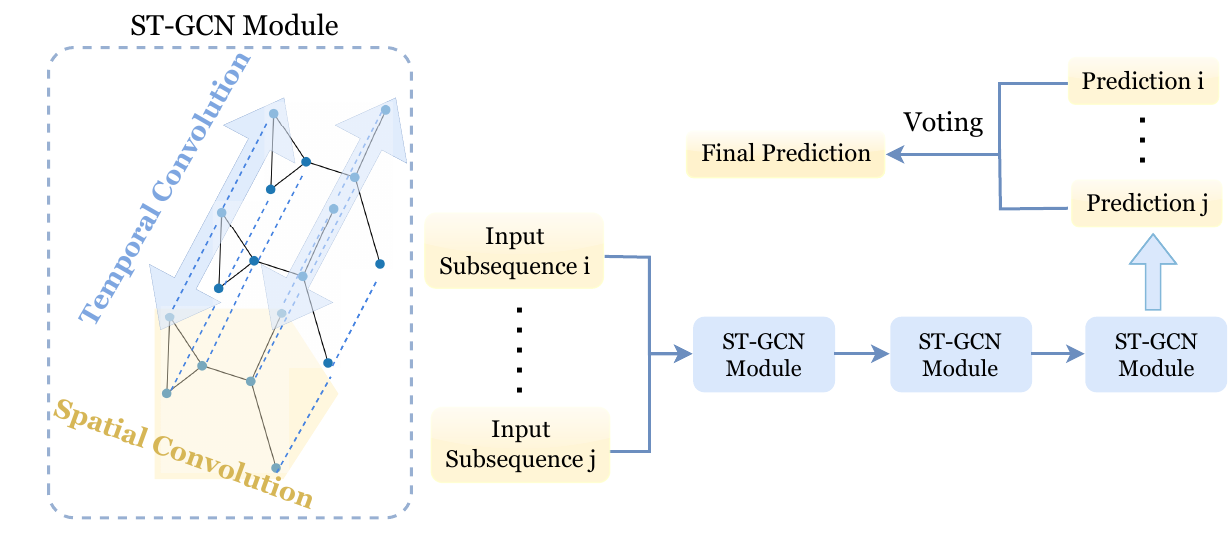}
    \caption{An illustration of the ST-GCN model architecture. Spatial graph convolution is first applied to the spatial graph at each time point. Then temporal convolution performs 1D convolution along the resulting features on each node. Multiple sub-sequences are randomly sampled from the whole time series as input graphs for training.}
    \label{fig:stgcn}
\end{figure*}

We adopt a popular model for fMRI classification, ST-GCN~\cite{gadgil2020spatio}, as the backbone architecture to extract graph representations from both spatial and temporal information. A graph convolution and a temporal convolution are performed in one ST-GCN module shown in Fig.~\ref{fig:stgcn}, following the details in~\cite{gadgil2020spatio}. The feature extractor includes three ST-GCN modules. The target head and the source head share the same architecture, which consists of one ST-GCN module and one fully-connected layer. 

To construct the graph, we treat brain regions parcellated by a brain atlas \cite{glasser2016multi} as the nodes and define edges using the functional connectivity between pairs of nodes measured by Pearson's correlation coefficient~\cite{bellec2017neuro}. We randomly sample sub-sequences from the whole fMRI time series to increase the size of training data by constructing multiple input graphs containing dynamic temporal information. 
For each time point in each node, a feature vector of dimension $C_i$ is learned. So for the $r$-th sub-sequence sample from the $n$-th subject, the input graph $X^{(n,r)}_i$ to the $i$-th layer has a dimension of $P\times L \times C_i$, where $P$ is the number of brain regions or parcels (nodes), $L$ is the length of the sampled sub-sequence, and $C_0=1$ for the initial input. 
In ST-GCN, a graph convolution~\cite{kipf2016semi}, applied to the spatial graph at time point $l$ in the $i$-th layer, can be expressed as follows.
\begin{equation}
    X^{(n, r, l)}_{i+1} = D^{-1/2}(A+I)D^{-1/2}X^{(n, r, l)}_i W_{C_i \times C_{i+1}}
\end{equation}
where $A$ is the adjacency matrix consisting of edge weights defined as Pearson's correlation coefficients, $I$ is the identity matrix, $D$ is a diagonal matrix such that $D_{ii}=\sum_j A_{ij} + 1$, and $W$ is a trainable weight matrix. 
We then apply 1D temporal convolution to the resulting sub-sequence of features on each node. A voting strategy is applied to combine predictions generated from different sub-sequences.
\subsection{Meta Knowledge Transfer}
We introduce a bi-level optimization strategy to perform gradient-based update of model parameters~\cite{finn2017model, liu2020meta}. The model first backpropagates the gradients through the target head only in several fast adaptation steps, and then backpropagates through the source head and feature extractor. Each step in a nested loop is summarized as follows: 

\textbf{Outer loop} ($M$ iterations): \texttt{Step 1}.  Initialize the target head and randomly sample target meta-training set $X_{\mathcal{T}_{tr}}$ and meta-validation set $X_{\mathcal{T}_{val}}$ from $X_\mathcal{T}$, where $X_{\mathcal{T}_{tr}} \bigcap X_{\mathcal{T}_{val}} = \emptyset$, $X_{\mathcal{T}_{tr}} \bigcup X_{\mathcal{T}_{val}} = X_\mathcal{T}$.

\textbf{\texttt{Step 2}. Inner loop} ($k$ update steps): Only target head parameters $\theta_t$ are updated using optimization objective $\mathcal{L_T}$ (see below) for the target task.  The  parameter $\alpha$ is the inner loop learning rate, and $\theta_t^j$ is the target head parameter at the $j$-th update step.
    \begin{equation}
        \theta_t^{j+1} = \theta_t^{j} - \alpha \nabla_{\theta_t^{j}} \mathcal{L_T} (h_\mathcal{T}(f(X_{\mathcal{T}_{tr}}; \phi^i); \theta_t^j))
    \end{equation}

\texttt{Step 3}: After the inner loop is finished, freeze the target head and update feature extractor parameters $\phi$ and source head parameters $\theta_s$. The target loss $\mathcal{L_T}$ and source loss $\mathcal{L_S}$ are defined in the following section. The parameter $\beta$ is the outer loop learning rate, and $\lambda$ is a scaling coefficient.\begin{equation}
    \begin{split}
    \{\theta_s^{i+1}, \phi^{i+1}\} =
         \{\theta_s^{i}, \phi^{i} \} - \beta (\nabla_{\theta_s^{i}, \phi^i} \mathcal{L_S} (h_\mathcal{S}(f(X_{\mathcal{S}}; \phi^i); \theta_s^i)) \\
         +  \nabla_{\phi^i} \lambda \mathcal{L_T} (h_\mathcal{T}(f(X_{\mathcal{T}_{val}}; \phi^i); \theta_t^k)))
    \end{split}
    \end{equation}
    The target head, source head and feature extractor are updated in an alternating fashion. The target head is first trained on $X_{\mathcal{T}_{tr}}$ in the inner loop. In the outer loop, the feature extractor and source head are trained to minimize the generalization error of the target head on an unseen set $X_{\mathcal{T}_{val}}$ as well as to minimize the source loss. In this way, the feature extractor encodes features beneficial for both domains and the source head extracts features from the source domain that enable generalization to the target domain.
\subsection{Contrastive Self-supervised Learning}
To further boost the generalizability of features, we apply a graph contrastive loss~\cite{you2020graph} to perform a self-supervised task on the source domain. We randomly sample sub-sequences $X^{(n,r_1)}$, $X^{(n,r_2)}$ $(r1 \neq r2)$ from the whole fMRI time series for subject $n$ as the input graph features~\cite{gadgil2020spatio}, which can be viewed as an augmentation of input graphs for ST-GCN. $X^{(n,r_1)}$ and $X^{(n,r_2)}$ should produce similar output graph features even though they contain different temporal information. The graph contrastive loss enforces similarity between graph features extracted from the same subject and dissimilarity between graph features extracted from different subjects~\cite{chen2020simple}, so that the model learns invariant functional activity patterns across different time points for the same subject and recognizes inter-subject variances. A cosine similarity is applied to measure the similarity in the latent graph feature space~\cite{you2020graph}.
\begin{equation}
        \mathcal{L_S} = \dfrac{1}{N} \sum_{n=1}^{N} -\log\frac{{\exp{( sim(\Tilde{X}_\mathcal{S}, n, n)/ \tau)}}}{\sum_{m=1, m\neq n}^{N} \exp{(sim(\Tilde{X}_\mathcal{S}, n, m)/ \tau)}}
\end{equation}
\begin{equation}
    sim(X, n, m) = \frac{(X^{(n, r_1)})^\top X^{(m, r_2)}}  { \|X^{(n, r_1)}\| \cdot \|X^{(m, r_2)}\|}
\end{equation}
where $\Tilde{X}_\mathcal{S} = h_\mathcal{S}(f(X_\mathcal{S}; \phi); \theta_s)$ is the generated graph representation, $\tau$ is a temperature hyper-parameter, and N is the total number of subjects in one training batch. By minimizing the graph contrastive loss on the source domain, the model produces consistent graph features for the same subject and divergent graph features across different subjects, which may be related to latent functional activities that reveal individual differences, and such features are generalizable across domains.

The optimization objective $\mathcal{L_T}$ of the target domain depends on the target task. In a classification task with class labels $Y_\mathcal{T}$, we adopt the Cross-Entropy loss. The total loss for the proposed strategy, MeTSK, is
\begin{equation}
\begin{aligned}
\mathcal{L}_{meta} &= \mathcal{L_S} + \lambda \mathcal{L_T} \\
\mathcal{L_T} &= -\sum_{\mathrm{classes}} Y_{\mathcal{T}}\log(h_\mathcal{T}(f(X_{\mathcal{T}}; \phi); \theta_t))
\end{aligned}
\end{equation}

\subsection{Domain Similarity}
To evaluate the generalization of learned features, we measure the distance between features extracted from different domains using domain similarity~\cite{cui2018large, oh2022understanding}. We first compute the Earth Mover's Distance (EMD)~\cite{1521516}, which is based on the solution to the Monge-Kantorovich problem~\cite{rachev1985monge}, to measure the cost of transferring features from the source to target domain. We define $\Bar{X}_\mathcal{S} = \operatorname{Flatten}(\dfrac{1}{N} \sum_{n=1}^{N} \Tilde{X}_\mathcal{S})$, $\Bar{X}_\mathcal{T} = \operatorname{Flatten}(\dfrac{1}{N} \sum_{n=1}^{N} \Tilde{X}_\mathcal{T})$ as the flattened vectors of the output graph features averaged over all subjects, and then define $B_s$ and $B_t$ as the set of bins in the histograms representing feature distribution in $\Bar{X}_\mathcal{S}$ and $\Bar{X}_\mathcal{T}$, respectively. Domain similarity (DS) is defined in Eq.~\ref{eq:ds} and Eq.~\ref{eq:emd}. A larger domain similarity indicates better transferability and generalizability from the source domain to the target domain because the amount of work needed to transform source features into target features is smaller.
\begin{equation}
\label{eq:ds}
    \mathrm{DS}  = \exp{(-\gamma \operatorname{EMD}(\Bar{X}_\mathcal{S}, \Bar{X}_\mathcal{T}))}
\end{equation}
\begin{equation}
\label{eq:emd}
\begin{aligned}
\operatorname{EMD}(\Bar{X}_\mathcal{S}, \Bar{X}_\mathcal{T}) & = \frac{\sum_{i=1}^{|B_s|} \sum_{j=1}^{|B_t|} f_{i, j} d_{i, j}}{\sum_{i=1}^{|B_s|} \sum_{j=1}^{|B_t|} f_{i, j}}, \\
s.t. \quad & f_{i j} \geq 0, \\
& \sum\limits_{j=1}^{|B_t|} f_{i j} \leq \dfrac{|\Bar{X}_\mathcal{S} \in B_s(i)|}{|\Bar{X}_\mathcal{S}|}, \\
& \sum\limits_{i=1}^{|B_s|} f_{i j} \leq \dfrac{|\Bar{X}_\mathcal{T} \in B_t(j)|}{|\Bar{X}_\mathcal{T}|}, \\
& \sum\limits_{i=1}^{|B_s|} \sum\limits_{j=1}^{|B_t|} f_{i j}  = 1
\end{aligned}
\end{equation}
where $B_s(i)$ is the i-th bin of the histogram and $|B_s|$ is the total number of bins, $|\Bar{X}_\mathcal{S} \in B_s(i)|$ is the number of features in $B_s(i)$, $|\Bar{X}_\mathcal{S}|$ is the total number of features, $d_{i,j}$ is the Euclidean distance between the averaged features in $B_s(i)$ and $B_t(j)$,  $f_{i,j}$ is the optimal flow for transforming $B_s(i)$ into $B_t(j)$ that minimizes the EMD. Following the setting in~\cite{cui2018large}, we set
$\gamma=0.01$. 

\section{Datasets}
In this section, we introduce the datasets used to build the foundation model. The HCP~\cite{van2013wu} and ADHD~\cite{bellec2017neuro} datasets described below are used during the upstream training of the foundation model, the ABIDE dataset~\cite{craddock2013neuro} is used in the ablation study of the proposed MeTSK strategy. We then introduce the PTE dataset that is used for evaluation of downstream performance.
\subsection{Foundation Model Datasets}
\textbf{HCP dataset: }The healthy control data for the foundation model is drawn from the  Human Connectome Project (HCP) S1200 dataset~\cite{van2013wu}. 
The HCP database includes 1,096 young adult (ages 22-35) subjects with resting-state-fMRI data collected at a total of 1200 time-points per session. The preprocessing of fMRI follows the minimal preprocessing procedure in~\cite{gadgil2020spatio, glasser2013minimal}. 
Finally, the brain was parcellated into $116$ Regions of Interest (ROIs) using the Automated Anatomical Labeling (AAL) atlas in~\cite{tzourio2002automated}. 
The AAL atlas was defined based on brain anatomy. It divides the brain into 116 regions, including 90 cerebrum regions and 26 cerebellum regions.
These 116 regions form the nodes of our graph. 
The fMRI data were reduced to a single time-series per node by averaging across each ROI.

\textbf{ADHD-Peking: }The Attention-Deficit/Hyperactivity Disorder (ADHD-200) consortium data from the Peking site~\cite{bellec2017neuro} includes 245 subjects in total, with 102 ADHD subjects and 143 Typically Developed Controls (TDC). 
To investigate the scenario where clinical data is scarce, we use only the subset of the larger ADHD database that was collected from the Peking site. 
We use the preprocessed data released on (\url{http://preprocessed-connectomes-project.org/adhd200/}). 
During preprocessing, the initial steps involve discarding the first four time points, followed by slice time and motion correction. The data is then registered to the Montreal Neurological Institute (MNI) space, processed with a bandpass filter (0.009Hz - 0.08Hz), and smoothed using a 6 mm Full Width at Half Maximum (FWHM) Gaussian filter.
The fMRI data consisted of $231$ time points after preprocessing. As a final step, the ADHD-Peking data were re-registered from MNI space to the same AAL atlas as for the HCP subjects, and the average time-series computed for each ROI.

\textbf{ABIDE-UM: }The Autism Brain Imaging Data Exchange I (ABIDE I)~\cite{craddock2013neuro} collects resting-state fMRI from 17 international sites. Similar to the ADHD dataset, we use only the subset of data from the UM site, which includes 66 subjects with Autism Spectrum Disorder (ASD) and 74 TDCs (113 males and 27 females aged between 8-29). 
We downloaded the data from~\url{http://preprocessed-connectomes-project.org/abide/}, where data was pre-processed using the C-PAC pre-processing pipeline~\cite{craddock2013neuro}.
The fMRI data underwent several preprocessing steps: slice time correction, motion correction, and voxel intensity normalization.  
The data was then band-pass filtered (0.01–0.1 Hz) and spatially registered to the MNI152 template space using a nonlinear method. All fMRIs have 296 time points. 
As a final step, the ABIDE-UM data were re-registered from MNI space to the same AAL atlas as for the HCP subjects, and the average time-series computed for each ROI. 

\subsection{Downstream Clinical PTE Dataset}
We use the Maryland TBI MagNeTs dataset~\cite{gullapalli2011investigation} for downstream performance evaluation. All subjects suffered a traumatic brain injury. Of these we used acute-phase (within 10 days of injury) resting-state fMRI from 36 subjects who went on to develop PTE and 36 who did not~\cite{gullapalli2011investigation,zhou2012default}. The dataset was collected as a part of a prospective study that includes longitudinal imaging and behavioral data from TBI patients with Glasgow Coma Scores (GCS) in the range of 3-15 (mild to severe TBI). The individual or group-wise GCS, injury mechanisms, and clinical information is not shared. The fMRI data are available to download from FITBIR (\url{https://fitbir.nih.gov}). In this study, we used fMRI data acquired within 10 days after injury, and seizure information was recorded using follow-up appointment questionnaires. Exclusion criteria included a history of white matter disease or neurodegenerative disorders, including multiple sclerosis, Huntington's disease, Alzheimer's disease, Pick's disease, and a history of stroke or brain tumors. 
The imaging was performed on a 3T 
Siemens TIM Trio scanner (Siemens Medical Solutions, Erlangen, Germany) using a 12-channel receiver-only head coil. The age range for the epilepsy group was 19-65 years (yrs) and 18-70 yrs for the non-epilepsy group.

Pre-processing of the MagNeTs rs-fMRI data was performed using the BrainSuite fMRI Pipeline (BFP) (\url{https://brainsuite.org}). 
BFP is a software workflow that processes fMRI and T1-weighted MR data using a combination of software that includes BrainSuite, 
AFNI, FSL, and MATLAB scripts to produce processed fMRI data represented in a common grayordinate system that contains both cortical surface vertices and subcortical volume voxels \cite{glasser2013minimal}. As described above, the pre-processed data were then mapped to the same AAL atlas as used with the other datasets. Regional time-series were then generated for each of the 116 parcels by averaging over the corresponding region of interest. 

\section{Experiments and Results}
\subsection{Upstream Results}
We first trained the foundation model using the proposed MeTSK strategy on the HCP data (healthy controls) and ADHD-Peking data (clinical data). 
To investigate the effectiveness of MeTSK, we designed an experiment for an upstream task that performs ADHD v.s. TDC classification. We evaluate different strategies and compare their effectiveness in enhancing the generalization from a healthy dataset to a clinical dataset.

For comparison, we designed (i) a baseline model using a ST-GCN with a supervised task directly trained on the ADHD-Peking data (Baseline),(ii) a ST-GCN model fine-tuned on ADHD-Peking data after pre-training on HCP data (FT), (iii) a model performing multi-task learning on HCP data and ADHD-Peking data simultaneously (MTL), and (iv) the proposed strategy, MeTSK. 
We incorporated MTL and FT methods for comparison in order to investigate whether MeTSK is superior to traditional approaches in terms of generalization to ADHD data. 
For the MTL implementation, we simply remove the inner loop in MeTSK and use all the training data to update the target head. Both heads and the feature extractor are updated simultaneously in one loop. 
We compared several baseline methods: a Linear Support Vector Machine (SVM), a Random Forest Classifier (RF), a Multi-Layer Perceptron (MLP) consisting of three linear layers, an LSTM model for fMRI analysis~\cite{gadgil2020spatio}, and a model combining a transformer and graph neural network (STAGIN)~\cite{kim2021learning}. For the SVM, RF, and MLP, the inputs are flattened functional connectivity features, calculated using the Pearson's correlation coefficient between fMRI time-series across pairs of brain regions defined in the AAL atlas. LSTM and STAGIN, on the other hand, utilize raw fMRI time-series as their input.

\begin{table*}[!t]
\centering
\caption{A comparison of mean AUCs and ACCs of 5-fold cross-validation on ADHD data using different methods:
baseline, fine-tuning, multi-task learning, the proposed strategy MeTSK, and other baseline methods.}
\label{tab:res}
\begin{tabular}{l|cc|cc}
\toprule[1pt]
\hline
Method & HCP & ADHD-Peking & AUC & ACC\\
\hline
SVM & \ding{55} & \ding{51} & $0.6182 \pm 0.0351$ & $0.6086 \pm 0.0412$ \\
RF& \ding{55} & \ding{51} &$0.6117 \pm 0.0503$ & $0.6102 \pm 0.0564$\\
MLP & \ding{55} & \ding{51} & $0.6203 \pm 0.0468$ & $0.6092 \pm 0.0507$\\
LSTM~\cite{gadgil2020spatio} & \ding{55} & \ding{51} & $0.5913 \pm 0.0510$ & $0.5652 \pm 0.0539$\\
STAGIN~\cite{kim2021learning} & \ding{55} & \ding{51} & $0.5638 \pm 0.0468$ & $0.5279 \pm 0.0511$\\
Baseline (ST-GCN) & \ding{55} & \ding{51}  & $0.6215 \pm 0.0435$  & $0.6171 \pm 0.0556$\\
FT & \ding{51} & \ding{51} & $0.6243\pm0.0483$ & $0.6367 \pm 0.0501$\\
MTL & \ding{51} & \ding{51} & $0.6518 \pm 0.0428$ & $0.6316 \pm 0.0513$ \\
\textbf{MeTSK (ours)} & \ding{51} & \ding{51} & $ \mathbf{0.6981 \pm 0.0409}$ & $\mathbf{0.6775 \pm 0.0443}$ \\
\bottomrule
\end{tabular}
\end{table*}

We use 5-fold cross-validation to split training/testing sets on ADHD-Peking data and use all HCP data for training. For meta-learning, the ADHD training set in each fold is further divided into a meta-training set $X_{\mathcal{T}_{tr}}$ of $157$ subjects and a meta-validation set $X_{\mathcal{T}_{val}}$ of $39$ subjects. Model performance is evaluated on the test ADHD data set using the average area-under-the-ROC-curve (AUC) and classification accuracy (ACC) as evaluation metrics as shown in Table~\ref{tab:res}. MeTSK achieved the best mean AUC of $0.6981$, which is a significant improvement compared to the baseline model trained only on ADHD data. MeTSK also surpassed the performance of fine-tuning and multi-task learning, providing evidence for overcoming limited generalization. 
The results from upstream training demonstrate that the MeTSK strategy possesses a clear capability to enhance generalization from healthy data to clinical data.

\subsection{Downstream Results on PTE Dataset}
For the downstream application we performed zero-shot evaluation on the PTE dataset. This involved initially extracting features from the PTE dataset using the pre-trained foundation model without any further fine-tuning. These extracted features are ``zero-shot'' features, as they are generated directly from the model trained on different datasets. Subsequently, we input these zero-shot features into a classifier to differentiate between PTE and non-PTE subjects, thereby assessing the model's ability to generalize and apply learned patterns to the downstream clinical applications. 

Training a foundation model with only self-supervised learning is a typical approach. 
To compare different pre-training strategies for the foundation model, we also pre-trained a ST-GCN model on both HCP and ADHD-Peking datasets using only the proposed contrastive self-supervised learning (SSL). From this pre-trained SSL model, we again generated zero-shot features for PTE data. 
We also compared our proposed foundation model to a large pre-trained fMRI model, as detailed in~\cite{thomas2022self}. This model involves pre-training a Generative Pretrained Transformer (GPT)~\cite{radford2019language} on extensive datasets comprising 11,980 fMRI runs from 1,726 individuals across 34 datasets. During pre-training, the GPT model performs a self-supervised task to predict the next masked time point in the fMRI time-series. Their pre-trained model is publicly available at \url{https://github.com/athms/learning-from-brains}. We directly applied their pre-trained model to generate zero-shot PTE features. 

Finally, we compare the zero-shot features generated from different foundation models with functional connectivity features extracted from raw fMRI data. We employed the same machine learning classifiers as used in the upstream experiments, including a linear SVM, RF, and MLP. The same 5-fold cross-validation was applied and AUCs for PTE v.s. non-PTE classification were computed. 

The zero-shot features generated by the foundation model pre-trained using the MeTSK strategy achieved the best performance among all features in every classifier, as shown in Table \ref{tab:pte}, indicating superior generalization of the foundation model on the heterogeneous PTE dataset. The zero-shot features generated by the SSL model also achieved better performance than functional connectivity features, owing to the generalizable knowledge learned from upstream datasets. However, ~\cite{thomas2022self} achieved the worst performance, possibly because this pre-trained model needs further fine-tuning to boost its optimal performance. 
Notably, the best performance achieved by Linear SVM suggests that these zero-shot features are linearly separable. 
This outcome not only demonstrates MeTSK’s ability to produce discriminative features for an unseen dataset like PTE but also highlights its potential in enhancing feature learning for clinical diagnostic purposes. 

\begin{figure*}[!t]
    \centering
\includegraphics[width=0.75\textwidth]{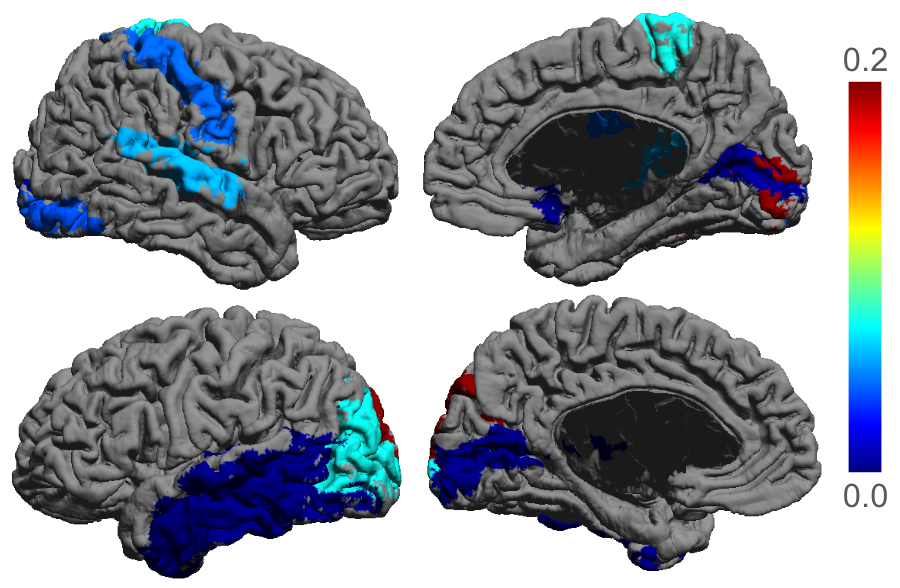}
    \caption{Feature importance map of zero-shot PTE features shown as color-coded ROIs overlaid on the AAL atlas.}
    \label{fig:pte_feat}
\end{figure*}

To gain further insights and improve the interpretability of the zero-shot PTE features from MeTSK, we computed a feature importance map derived from the positive SVM coefficients. In a linear SVM, each feature in each ROI is assigned a coefficient, indicating its significance in the decision-making process of the model. The higher the absolute value of a coefficient, the more impact that feature has on the model's predictions. We derived the coefficients for features of each ROI from the trained SVM and visualized these coefficients in the form of a feature importance map overlaid on the brain, which is shown in Fig.~\ref{fig:pte_feat}. Through observing the feature importance map, we can identify and interpret the most significant brain regions for PTE classification, which are mainly located in the temporal lobe. Given that epilepsy most commonly occurs in the temporal lobe, these significant brain regions identified from the zero-shot features offers potentially meaningful insights into the prediction of PTE. Interestingly, the other areas of high feature importance are in primary sensory (visual and somatomotor) regions. 
\begin{table*}[!ht]
\centering
\caption{Downstream results using 5-fold cross-validation: Mean and std of AUCs for PTE classification using zero-shot features generated from different foundation models as well as functional connectivity features.}
    \begin{tabular}{l c c c c}
    \toprule
    \hline
    & \multicolumn{3}{c}{Zero-shot Features} & \multirow{2}{*}{Connectivity Features} \\
    \cmidrule(lr){2-4} 
        & MeTSK  & SSL & \cite{thomas2022self} &  \\
       \hline
SVM & $\mathbf{0.6415 \pm 0.0312}$ & $0.5972 \pm 0.0492$ &$0.5369 \pm 0.0451$ &  $0.5697 \pm 0.0477$ \\
RF & $0.5392 \pm 0.0553$ & $0.5253 \pm 0.0486$ & $0.4814 \pm 0.0664$ & $0.5081 \pm 0.0612$ \\
MLP & $0.5813 \pm 0.0504$ & $0.5216 \pm 0.0329$ & $0.5278 \pm 0.0643$ & $0.5111 \pm 0.0402$\\
\hline
    \bottomrule
    \end{tabular}
\label{tab:pte}
\end{table*}

\subsection{Implementation Details}
\textbf{Upstream: }To optimize model performance, we follow the training setting in~\cite{gadgil2020spatio} for the ST-GCN model. 
We generate one meta-training batch by randomly selecting an equal number of samples from each class. The batch size is 32, both for the meta-training and the meta-validation set. We use an Adam optimizer~\cite{kingma2014adam} with learning rate $\beta=0.001$ in the outer loop, and an SGD optimizer~\cite{ketkar2017stochastic} with learning rate $\alpha=0.01$ in the inner loop. The number of inner loop update steps is $25$. 
We set the hyper-parameter $\lambda=30$ and the temperature parameter $\tau=30$ to adjust the scale of losses following~\cite{liu2020meta, you2020graph}. 
Since contrastive loss converges slowly~\cite{jaiswal2020survey}, a warm-up phase is applied to train the model only on HCP data using the graph contrastive loss for the first half of total training steps. 

\textbf{Downstream: }We use the pre-trained feature extractor for generating zero-shot features. The generated features are graph-level representations, having a two dimensional feature matrix at each node (brain region). We averaged the features along the first dimension and applied Pinciple Component Analysis (PCA) to reduce the dimensionality before feeding the features into classifiers. The MLP used in the experiments consists of three linear layers, with hidden dimensions of 32, 16, 16. The SSL model trained on both HCP and ADHD-Peking data used the same contrastive loss. In our comparative analysis with another foundation model for fMRI~\cite{thomas2022self}, we flatten the brain signals at each time-point and input the whole time-series without masking into the pre-trained GPT model. This generates a feature embedding for each time-point, which is then averaged within each time-point and fed into classifiers. We follow the other detailed settings of the pre-trained GPT model in~\cite{thomas2022self}. We ran $100$ iterations 
of stratified cross-validation on the PTE data for each method.

\section{Ablation Study and Generalization Analysis} 

\subsection{Experiments on ABIDE-UM}
\begin{figure*}[!t]
    \centering
\includegraphics[scale=.55]{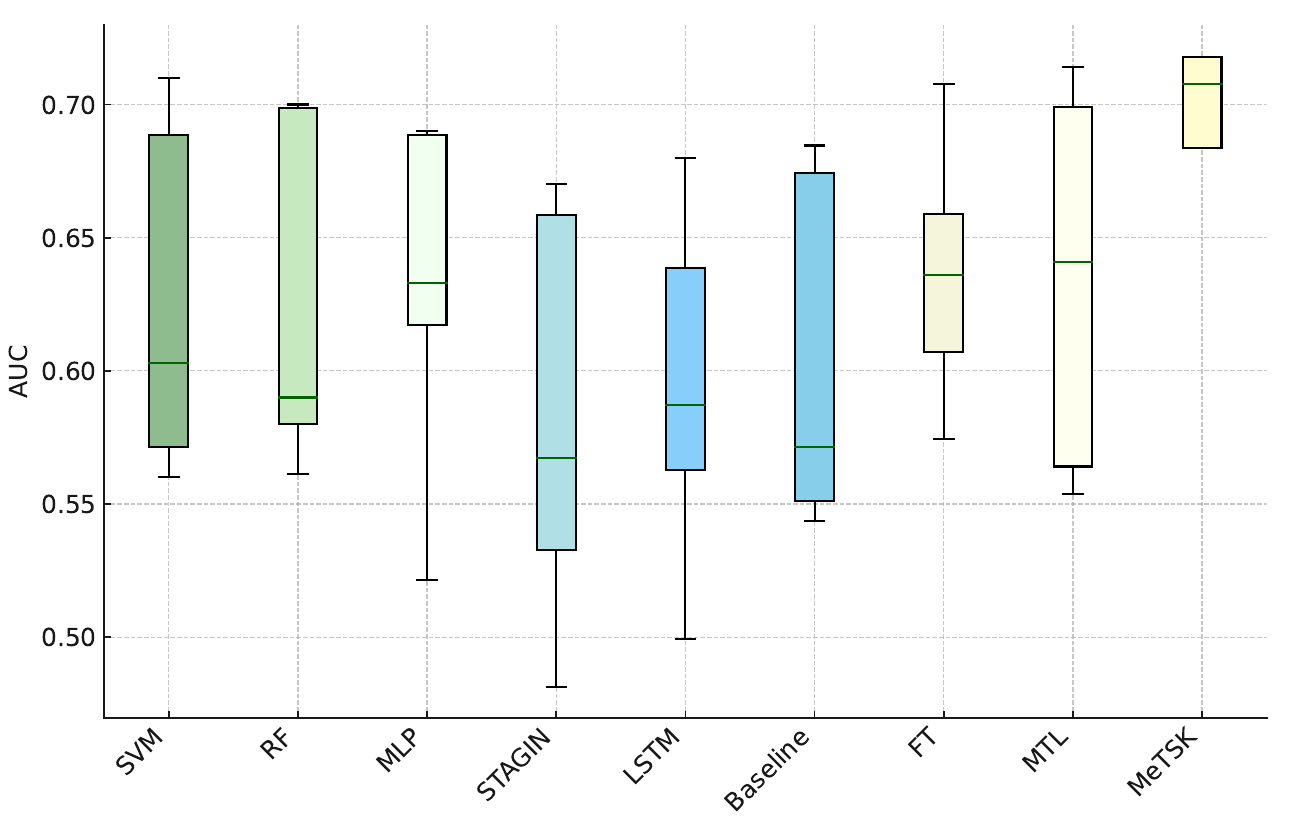}
\caption{AUCs of 5-fold cross-validation on the ABIDE-UM dataset: a comparison of baseline (ST-GCN), fine-tuning (FT), multi-task learning (MTL), and other baseline methods.}
    \label{fig:boxplot}
\end{figure*}

To investigate the robustness of our proposed pre-training strategy, MeTSK, across various clinical datasets, we also conducted experiments using the ABIDE-UM dataset as the target clinical dataset during upstream training. The same methods were compared and same experimental settings were applied to the ABIDE-UM data as for ADHD-Peking. We performed ASD v.s. TDC binary classification using the same 5-fold cross-validation. 
As shown in Fig.~\ref{fig:boxplot}, the performance on the ABIDE-UM dataset aligns with our findings for the ADHD-Peking dataset, with MeTSK consistently achieving the highest mean AUC among all compared methods. The results on ABIDE-UM illustrate MeTSK's applicability in different clinical datasets. When the downstream clinical task shares more similarities with ASD features or other clinical features, the training strategy of the foundation model can be adjusted to leverage different clinical features, demonstrating the flexibility of MeTSK in accommodating varying clinical datasets.

\subsection{Ablation Study of MeTSK}
We examine the individual contributions of self-supervised learning and meta-learning to the model performance during upstream training on both target clinical datasets (ADHD-Peking, ABIDE-UM) in this section. To explore the effect of meta-learning, we designed an experiment using only the target (clinical) dataset in meta-learning (MeL). This approach involves removing the source head and the source loss during bi-level optimization. The target head is first trained on the ADHD/ABIDE meta-training set in the inner loop, followed by feature extractor learning to generalize on a held-out validation set in the outer loop. Our results, as shown in the last two rows of Table~\ref{tab:ablation}, reveal that the mean AUC improved from $0.6215$ to $0.6562$ for ADHD classification, and from $0.6085$ to $0.6675$ for ASD classification without source domain knowledge. This finding is consistent with an increased generalization achieved by meta-learning on the clinical datasets.

Furthermore, to assess the contribution of self-supervised learning, we compared the impact of using a self-supervised task versus a sex classification task on the HCP dataset. Fine-tuning, multi-task learning, and MeTSK were implemented using sex classification (female vs male) as the source task. The same 5-fold cross-validation method was applied to compare the average AUC. As detailed in Table~\ref{tab:ablation}, all three methods: FT, MTL, and MeTSK, showed a degraded performance when transferring knowledge from the sex classification task. This suggests that the sex-related features of the brain may be less relevant to ADHD/ASD classification, negatively affecting the model's performance.
\begin{table*}[!t]
\centering
\caption{Ablation study on ADHD-Peking and ABIDE-UM dataset. The FT, MTL, and MeTSK methods are compared for two cases - transferring features from (i) a self-supervised source task and (ii) a sex classification source task, respectively. The last two rows are models trained only on target clinical data: a meta-learning model without source task and a baseline model.}
\begin{tabular}{l c c | c c}
\toprule[1pt]
 Dataset & \multicolumn{2}{c}{ADHD-Peking} & \multicolumn{2}{c}{ABIDE-UM}\\
 \cline{1-5} 
Source Task & Self-supervision & Sex Classification & Self-supervision & Sex Classification\\
\hline
 FT
 & $0.6213\pm0.0483$ & $0.6150 \pm 0.0497$ & $0.6368 \pm 0.0454$ & $0.6071 \pm 0.0742$ \\
MTL & $0.6518 \pm 0.0428$ & $0.6377 \pm 0.0512$ & $0.6345 \pm 0.0663$ & $0.6240 \pm 0.0711$ \\
\textbf{MeTSK}& $ 0.6981 \pm 0.0409$ & $0.6732 \pm 0.0579$ & $0.6967 \pm 0.0568$ & $0.6786 \pm 0.0749$\\
\hline
MeL &\multicolumn{2}{c}{ $0.6562 \pm 0.0489$} & \multicolumn{2}{c}{$0.6675\pm0.0505$} \\
Baseline &\multicolumn{2}{c}{ $0.6215 \pm 0.0435$} & \multicolumn{2}{c}{$0.6051\pm0.0615$} \\
\bottomrule[1pt]
\end{tabular}
    \label{tab:ablation}
\end{table*}

\subsection{Generalization Analysis Using Domain Similarity}
\begin{figure*}[!t]
    \centering
\includegraphics[scale=.75]{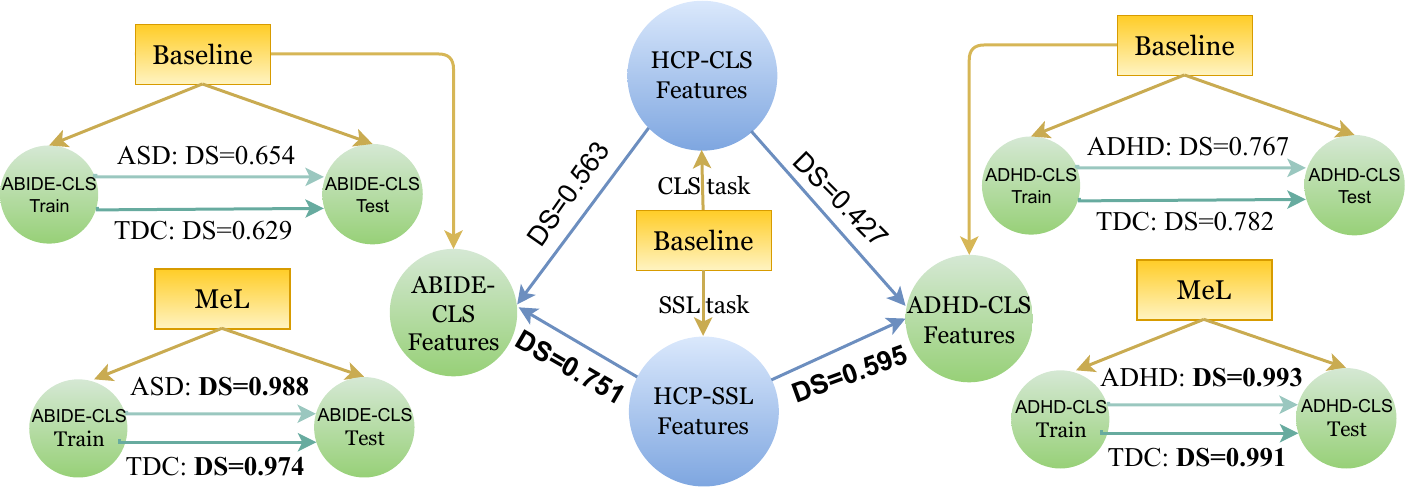}
    \caption{A comparison of the domain similarity between HCP self-supervised features (HCP-SSL, from Baseline ST-GCN trained on HCP data with a self-supervised task) and ADHD/ASD classification features (ADHD-CLS, ABIDE-CLS, from Baseline trained using all ADHD/ABIDE data), the domain similarity between HCP sex classification features (HCP-CLS, from Baseline trained on HCP data with a sex classification task) and ADHD-CLS/ABIDE-CLS, the intra-class (ADHD; TDC and ASD; TDC) domain similarities between training and testing set of ADHD/ASD data from Baseline and MeL (a meta-learning model trained only on target data), respectively.}
    \label{fig:ds}
\end{figure*}

To further investigate the generalization enabled by MeTSK, domain similarity was computed to evaluate the generalizability from control data (source) to clinical data (target) as well as from the training set to the testing set of target data. We conducted domain similarity analysis on both ADHD-Peking and ABIDE-UM datasets to further validate the robustness and versatility of MeTSK. Fig.~\ref{fig:ds} illustrates that the self-supervised source features have a higher similarity with the target features, indicating better inter-domain generalizability and thus improved performance on the target classification task. Moreover, compared to the baseline, both intra-ADHD-class/intra-ASD-class and intra-TDC-class domain similarities between the training and testing sets of ADHD/ABIDE data are increased by MeL. This enhancement provides evidence to explain the improved classification performance on training with only target data achieved by meta-learning. By applying meta-learning, not only the inter-domain generalization of features is boosted, but also the effect of heterogeneous data within the same domain is alleviated.

\section{Discussion and Conclusion}
Our proposed strategy opens up new possibilities for enabling data-efficient generalization to downstream applications and handling extremely heterogeneous and scarce datasets that eluded traditional deep-learning approaches. According to~\cite{kumar2022fine}, fine-tuning can distort good pre-trained features and degrade downstream performance under large distribution shifts. So unlike the common fine-tuning methods used in other foundation model approaches for fMRI analysis~\cite{ortega2023brainlm, thomas2022self}, we explored zero-shot features and linear probing for downstream adaptation, which achieved superior performance on the challenging PTE prediction task. Despite the  improvements achieved, 
exciting future work still remains to be explored. We trained the foundation model on one healthy control dataset and one clinical dataset, an approach that is sensitive to the cost of data collection and expert annotation. Without these constraints, multiple datasets could be combined to learn generalizable functional activity patterns from a diverse span of subjects and clinical conditions.

To tackle the heterogeneity and scarcity of fMRI data, 
we propose a novel training strategy for developing a foundation model by learning from both clinical and healthy fMRI data. We integrate meta-learning with self-supervised learning to improve the generalization from normal features to clinical features during upstream training, and thus enhance the generalization to other unseen clinical features in a downstream task for predicting post-traumatic epilepsy. 
Specifically, we perform a self-supervised task on the healthy control dataset and apply meta-learning to transfer self-supervised knowledge to the clinical dataset. 
To explore the generalizability of the foundation model to a post-traumatic epilepsy (PTE) dataset, we compared zero-shot features generated by different foundation models for PTE classification. The features from MeTSK demonstrated the best performance. Additionally, the interpretation of the zero-shot PTE features may contribute to our understanding of PTE, offering insights into the identification of PTE via functional brain activity patterns in different brain regions. 
To summarize, the improved generalization of our foundation model in predicting PTE is attributed to: (i) the application of meta-learning, which bolsters the model's generalization to clinical features, and (ii) the use of self-supervised features that are inherently more task-agnostic and more generalizable.

\section*{Declaration of Competing Interest}
The authors declare no competing interests.
\section*{Acknowledgments}
This work is supported by NIH grants: R01EB026299, R01NS074980 and DoD grants: W81XWH181061, HT94252310149.

\bibliographystyle{model2-names.bst}\biboptions{authoryear}
\bibliography{refs, pte_refs, gpt_refs}



\end{document}